\documentclass[runningheads]{llncs}
\usepackage{amssymb}
\usepackage{graphicx}
\usepackage{verbatim}  
\usepackage{url}
\usepackage{enumerate}
\urldef{\mailsa}\path|xiaowen.zhang@csi.cuny.edu|
\urldef{\mailsb}\path|cchum@gc.cuny.edu|
\newcommand{\keywords}[1]{\par\addvspace\baselineskip
\noindent\keywordname\enspace\ignorespaces#1}

\begin{document}

\mainmatter  

\title{Improved Latin Square based Secret Sharing Scheme}

\titlerunning{Improved Latin Square based Secret Sharing Scheme}

\author{Chi Sing Chum$\, ^{1}$ and Xiaowen Zhang$\, ^{1,2}$}

\authorrunning{Improved Latin Square based Secret Sharing Scheme}

\institute{$^1$~Computer Science Dept., Graduate Center / CUNY,\\
365 Fifth Ave., New York, NY 10016, U.S.A.\\
$^2$~Computer Science Dept., College of Staten Island / CUNY,\\
2800 Victory Blvd, Staten Island, NY 10314, U.S.A.}

\maketitle

\begin{abstract}

This paper first reviews some basic properties of cryptographic hash
function, secret sharing scheme, and Latin square. Then we discuss
why Latin square or its critical set is a good choice for secret
representation and its relationship with secret sharing scheme.
Further we enumerate the limitations of Latin square in a secret
sharing scheme. Finally we propose how to apply cryptographic hash
functions, herding attack technique to a Latin square based secret
sharing scheme to overcome these limitations.

\end{abstract}

\section{Introduction}

How to set up an effective procedure to keep a secret is important.
However, how to represent the secret is equally important. If we can
discover the secret by exhaustive search, then we can bypass the
secret sharing scheme, no matter how good it is. Also, it would be
efficient to keep the secret short, and difficult to discover at the
same time. Latin square is a good candidate in a secret sharing
scheme. We can use a Latin square to represent the secret, because
of the huge number of different Latin squares for a reasonably large
order. For example, there are about $10^{37}$ different Latin
squares of order 10. This makes outsiders difficult to discover the
secret without any knowledge due to the tremendous possibilities. We
can even improve the efficiency by distributing the shares of the
critical set, instead of the full Latin square, to the participants.
Whenever any group of the participants join together to form any
critical set, the original Latin square and hence the secret can be
recovered.\\

There are Latin square based secret sharing schemes in the
literature. Cooper, Donovan, Seberry \cite{CDS94} used critical sets
of Latin square in the design of secret sharing schemes. Their
schemes are not perfect because each share of a participant is a
component of a critical set. Therefore each share contains partial
information of the secret. Chaudhry and Seberry \cite{CS96} had
another secret sharing scheme based on critical sets of Room
squares. This scheme is not perfect, either. Distributing shares of
a critical set is fast and efficient. However it's not easy to
reconstruct the full Latin square, which is the shared secret, from
the critical set. Chaudhry, Ghodosi, Seberry \cite{CGS98} proposed a
perfect secret sharing scheme from Room squares, but the scheme is
not flexible, nor ideal. Each participant needs to have different
share for different authorized set he/she belongs to. It's not
flexible to set up a verifiable, or proactive secret sharing scheme
by just using Latin square or its critical sets, because it's hard
to verify a critical set for a large order Latin square. \\

In order to conquer the aforementioned limitations of Latin square
in a secret sharing scheme, we propose to apply cryptographic hash
functions, herding attack technique to Latin square based secret
sharing schemes. We can use hash function to store a partial Latin
square in a hash, such partial Latin square is easily extended to
the full Latin square. Then we set up a Latin square based ideal
perfect $(t+1, n)$ threshold scheme, which utilizes the herding hash
function and Nostradamus attack technique to iterative hash
functions. Finally we use two hash functions to set up a verifiable
secret sharing scheme, the method applies to any general secret
sharing schemes, including Latin square based schemes. The security
of our newly proposed schemes are dramatically improved. \\

In this section we review some basic properties of cryptographic
hash functions, herding attacks, and secret sharing schemes. In
Section 2 we discuss Latin square, partial Latin square, critical
set, and other concepts of Latin square. Section 3 presents
applications of critical set in secret sharing schemes. Section 4
discusses the limitations of Latin square in a secret sharing
scheme. In Section 5 we propose the applications of hash functions
to Latin square based secret sharing schemes with three examples.
Section 5 concludes the paper and summarizes the advantages of the
schemes we have designed.

\subsection{Cryptographic hash functions}

A cryptographic hash function \cite{STI05,TW06} takes an input
string of arbitrary length and generates an output string of fixed
length, which is called message digest, or hash value, or just
``hash''. Hash functions have many applications in information
security area, such as digital signatures, message authentication
codes, and authentication protocols. The following are common
properties that a well designed cryptographic hash function should
have.

\begin{enumerate}[1)]

\item Given an input string of arbitrary length, the output string will
be of fixed length. The output is usually called a hash value or
message digest.

\item For all practical purposes, given any message $x$, the message digest
$h(x)$ can be calculated very quickly.

\item Given a message digest $y$, it is computationally infeasible to
find $x$ such that $h(x) = y$. This, together with b), implies that
$h$ is a one way function, or preimage resistant.

\item Given an input and output pair $(x, y)$ for a hash function,
it should remain infeasible to find a second preimage $x'$ such that
$x \neq x'$ but $h(x) = h(x') = y$. This property is called second
preimage resistance.

\item It is infeasible to find two different inputs, $x$ and $x'$, that
produce the same output, i.e. $x \neq x'$ but $h(x) = h(x')$. This
property is called collision resistance.
\end{enumerate}

A hash function must have the flexibility to process messages of
arbitrary length. Most currently used hash functions, such as MD
family and SHA family, are built from iterations of a compression
function $C$ using Merkle-Damg{\aa}rd construction
\cite{DAM89,MER89}, they are also called \textbf{iterative hash
functions}. The process is as follows. (a) Pad the arbitrary length
message $M$ into multiple $v$-bit blocks: $m_1, m_2, \ldots, m_b$.
(b) Iterate the compression function $h_i = C(h_{i-1}, m_i)$, where
$i$ is from 1 to $b$ and $h_0$ is the initial value (or initial
vector) IV. (c) Output $h_b$ is the hash of the message $M$, i.e.,
$H(M) = h_b = C(h_{b-1}, m_b)$.\\

\subsection{Herding and Nostradamus attack}

Iterative hash functions are also vulnerable to herding and
Nostradamus attack. This attack also makes use of the fact that it
is not difficult to find intermediate hash values that can be
substituted for genuine blocks during iterative application of a
compression function and generate the same final hash value, $h$.
Kelsey and Kohno \cite{KK06} have a detailed analysis of this
attack. Stevens, Lenstra and Weger \cite{SLW07} applied the
technique to predict the winner of the 2008 US Presidential
Elections using a Sony PlayStation 3 in November 2007. They claimed
that they have correctly predicted the next US president, and
committed the hash of the result to the public. And the correct
prediction and the matching hash will be revealed after the
election. \\

The first step is to build a large set of intermediate hashes at the
first level: $h_{11}, h_{12}, \ldots, h_{1w}$. The second step is to
build a set of intermediate hashes at the second level:
$h_{21},h_{22}, \ldots, h_{2 w/2}$ so that the followings are
satisfied:\\

\noindent there exists a message $m_{11}$ such that
$C(h_{11},m_{11}) =h_{21}$ \\
there exists a message $m_{12}$ such that $C(h_{12},m_{12}) =
h_{21}$ \\
there exists a message $m_{13}$ such that $C(h_{13},m_{13}) =
h_{22}$ \\
there exists a message $m_{14}$ such that $C(h_{14},m_{14}) =h_{22}$
\\
$\ldots \ldots \ldots \ldots \ldots \ldots \ldots \ldots$
$\ldots \ldots \ldots \ldots \ldots \ldots \ldots \ldots$ \\

\begin{figure}
\centering
\includegraphics[height=5cm, width=8cm]{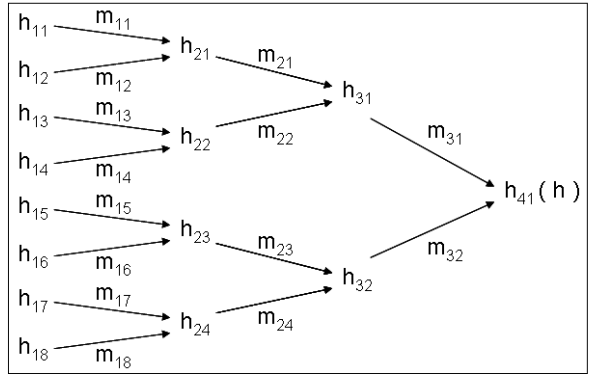}
\caption{A simplified diamond structure.} \label{fig:DIAM}
\end{figure}

By repeating this process, message blocks are linked so that each
intermediate hash at level 1 can reach the final hash, say $h$.
This is called the diamond structure (see Fig. \ref{fig:DIAM}). \\

We claim we can predict something happens in the future by
announcing this hash to the public. When the result is available, we
construct a message as follows:

$$M = (Prefix \| M^* \| Suffix),$$

\noindent where $Prefix$ contains the results that we claimed we
knew before it happens. $M^*$ is a block of message which can link
the $Prefix$ to one of the intermediate hash at level 1. $Suffix$ is
the rest of message blocks which linked the $M^*$ to the final hash.

\subsection{Secret sharing schemes}

A secret sharing scheme \cite{STI05,TW06} is a method to split and
distribute a secret among a group of participants, each of whom
receives a share of the secret. The secret can only be recovered
when the participants join together to combine their shares. \\

There are many practical applications of secret sharing schemes. For
example, they can be used to protect a private key from access by
outsiders. When we examine the problem of maintaining sensitive
information, we will consider two issues: \textbf{availability and
secrecy}. If only one person keeps the entire secret, then there is
a risk that the person might lose it or the person may not be
available when it is needed. We can solve the availability and
reliability issues by letting more than one person keep the same
secret. But the more people who can access the secret, the higher
the chance the secret will be leaked. A secret sharing scheme is
designed to solve these issues.\\

In 1979 Shamir \cite{SHA79} proposed the $(t+1, n)$ threshold
scheme, in which a secret is divided into pieces (shares) and
distributed among $n$ participants whereby any group of $t + 1$ or
more participants $(t \leq n -1)$ can recover the secret. Any group
of fewer than $t + 1$ cannot recover the secret. By sharing a secret
in this way the availability and reliability issues can be solved.\\

Shamir's scheme allows no partial information given out even up to
$t$ participants joined together \cite{STI05}. In other words, any
group of up to $t$ participants cannot gather more information about
the secret than any outsider. A secret sharing scheme with this
property is called a \textbf{perfect secret sharing scheme}. If the
shares and the secret come from the same domain, we call it an
\textbf{ideal secret sharing scheme}. In this case, the shares and
the secret have the same size.\\

Shamir's original sharing scheme assumes the dealer and all the
participants are honest. However, in reality, we need to consider
the situation that the dealer or some of the participants are
malicious. In this case, we need to set up a \textbf{verifiable
secret sharing scheme} so that the validity of a share of the
participants can be verified. In order to make this possible,
additional information is required for the participants to verify
their shares as consistent. Feldman's scheme \cite{FEL87} is a
simple verifiable secret sharing scheme that is based on Shamir's
scheme. It is based on the homomorphic properties of the
exponentiation function: $x^{a+b} = x^a \cdot x^b.$ \\

Many existing secret sharing schemes are subject to certain
limitations. One particular scheme is only applicable to one
specific access structure. If we want to apply one scheme to another
access structure, either it doesn't work or it's inefficient.
Although Ito, Saito, and Nishizeki \cite{ISN87} proved that any
general access structure can be realized by a secret sharing scheme,
but there is no guarantee that the scheme is efficient. Also, any
secret sharing scheme may not have all the desired properties such
as perfect, ideal, verifiable, and proactive. \\

\section{Latin square\label{Sec_LS}}


A Latin square of order $n$ is an array consists of $n$ rows and $n$
columns such that for any row and any column only one out of the $n$
symbols occurs exactly once. For simplicity, we usually use $0,
\ldots, n-1$ to represent the symbols so that each entry in a Latin
square can be represented as a triple $(i, j, k)$, where $0 \leq  i,
j, k \leq n-1$, and $i, j, k$ are the row, the column and the
symbol, respectively. For any order n, there exists a Latin square
of this order. The addition table of the additive group
$\mathbb{Z}/n\mathbb{Z}$ of integers mod n is an example
\cite{MM07}.

\subsection{Use a Latin square as a secret}

Suppose we use a Latin square to represent the secret and its order,
$n$, is made public. For an empty $n \times n$ array, there are $n!$
ways to fill out the first row. Now consider the second row. There
are $n-1$ choices for filling the `0'. There are $n-1$ or $n-2$
choices for filling the `1' depending on whether the `0' was filled
under the `1' in the first row or not. So there are at least $n-2$
choices for filling the `1'. We continue with `2', there are at
least $n-3$ choices. So, there are at least $(n-1)!$ ways to fill
out the second row. By similar argument, we can see there are at
least $n! (n-1)! (n-2)! \ldots 2!$ Latin squares of order $n$. This
is just a lower bound. For a reasonably large $n$, say $n > 10$,
there are many different Latin squares of this order. This
definitely makes an outsider very difficult to figure out the secret
itself without having any related knowledge.\\

The larger the order $n$ is, the larger the number of Latin squares
will be. For instance the number of Latin squares of order 10 and 11
are as follows \cite{McKW05,MM07}.

$L_{10} = 10! \times 9! \times 7,580,721,483,160,132,811,489,280;$

$L_{11} = 11! \times 10! \times
5,363,937,773,277,371,298,119,673,540,771,840.$

\noindent The number of Latin square of a given order is an open
problem. By now, the number of Latin squares of order 12 has not
been determined.

\subsection{Partial Latin square and extension of a partial Latin square}

A partial Latin square of order $n$ is an array that consists of $n$
rows and $n$ columns such that for any row and any column no symbol
occurs more than once and one or more cells(s) can be empty. I.e,
there exists one or more pair $(i, j)$ such that there is no symbol
in row $i$ and column $j$.\\

Some partial Latin squares can be extended to Latin squares of the
same order, while others cannot be. In the following example (see
Tab.~\ref{Tab_partialLS}), the partial Latin square on the left can
be extended into a Latin square in the middle. But the Latin square
on the right cannot be extended to a Latin square.

\begin{table}[ht]
\caption{Partial Latin square extendibility. \label{Tab_partialLS}}
  \begin{minipage}{1in} \centering
    \begin{tabular}{| c | c | c | c |}
        \hline
        $\; 0 \;$ &           & $\; 3 \;$ &            \\
        \hline
                  & $\; 2 \;$ &           &            \\
        \hline
                  &           & $\; 1 \;$ &            \\
        \hline
                  &           &           & $\; 3 \;$  \\
        \hline
    \end{tabular}
  \end{minipage}
  \hspace{0.5cm}
  \begin{minipage}{1in} \centering
    \begin{tabular}{| c | c | c | c |}
        \hline
        $\; 0 \;$ & $\; 1 \;$ & $\; 3 \;$ & $\; 2 \;$ \\
        \hline
            3     & $\; 2 \;$ &     0     &     1      \\
        \hline
            2     &     3     & $\; 1 \;$ &     0      \\
        \hline
            1     &     0     &     2     & $\; 3 \;$  \\
        \hline
    \end{tabular}
  \end{minipage}
  \hspace{0.5cm}
  \begin{minipage}{1in} \centering
    \begin{tabular}{| c | c | c | c |}
        \hline
        $\; 0 \;$ &           & $\; 3 \;$ & $\; 1 \;$ \\
        \hline
                  &           &           &            \\
        \hline
                  &           &           &            \\
        \hline
                  & $\; 2 \;$ &           &             \\
        \hline
    \end{tabular}
  \end{minipage}
\end{table}

In 1960, Trevor Evans conjectured that any partial Latin square of
order $n$ can be always extended to a full Latin square if the size
of the partial Latin square is up to $n-1$ \cite{EVA60}. Twenty
years later, this was proved to be true by Smetaniuk \cite{SME81}.
$n-1$ is the optimal number as we can see from the last table in
Tab.~\ref{Tab_partialLS}. \\

We define a partial Latin square as a Latin rectangle if the first
$m$ rows are all filled $(m < n)$ and the remaining $n-m$ rows are
all empty. A Latin rectangle can always be extended to a full Latin
square by adding row by row. This can be proved by Hall's condition
in prefect matching \cite{HALL35}. However, whether an arbitrary
partial Latin square can be extended to a full Latin square is an
NP-complete problem \cite{COL84}. Also, given a partial Latin
square, there may be different ways to extend it to different Latin
squares of the same order.

\subsection{Critical set and strong critical set}

A critical set of a Latin square is a partial Latin square which can
be extended to a full Latin square uniquely. In other words, there
is only one Latin square which contains the critical set. After
deletion of any entry of a critical set, the unique completion
property does not hold any more. For a given Latin square, there may
exist critical sets of different sizes. \\

By definition, we know we can recover the original Latin square from
one of its critical set and the completion is unique. However,
whether we can complete to a Latin square from a partial Latin
square is an NP-complete problem \cite{COL84}. That means the
recovery of the Latin square from one of its critical set may be
time-consuming. We really need some criteria to speed up the
process.\\

Donovan, Cooper, Nott and Seberry \cite{DCNS95} defined a strong
critical set. Let $L$ be a Latin square of order $n$ and $C$ one of
its critical set. Let $|C|$ be the size of $C$, the number of non
empty cells in $C$. If there is a sequence of partial Latin squares
$\{P_1, P_2, \ldots, P_m\}$ such that

\begin{enumerate}[1)]

\item $C = P_0 \subset P_1 \subset \ldots \subset Pm = L$, where $m = n^2 -
|C|$;

\item for any $i, 0 \leq i \leq m-1,  P_i \cup \{(r_i, c_i, k_i) \} = P_{i+1}$ and
$P_i \cup \{(r_i, c_i, k)\}$ is not a partial Latin square if $k
\neq k_i$.

\end{enumerate}

That means we start from the critical set $C$ and enter an entry one
at a time until we finish the extension to a full Latin square $L$.
When we get a new partial Latin square $P_{i+1}, \; 0 \leq i \leq m
- 1$ each time, there always exists a cell $(r_i, c_i)$ that can be
filled with only one symbol $k_i$. We call such critical set as a
strong critical set if it has the above properties. In other words,
the `force out' process makes a strong critical set to be extended
to a full Latin square easily.

\section{Application of critical set in secret sharing}

Cooper, Donovan, Seberry \cite{CDS94} proposed to form a collection
of critical sets of a Latin square, say $S$. Elements of $S$ are
distributed to participants. Any group of participants is an
authorized group if their shares pooled together is one of the
critical sets forming $S$.\\

\noindent (1) For example:  A $(2, 3)$ threshold scheme is shown in
Tab.~\ref{Tab_23ThresholdScheme}.

\begin{table}[ht]
\caption{A $(2, 3)$ threshold secret sharing
scheme.\label{Tab_23ThresholdScheme}}
  \begin{minipage}{1in} \centering
    \begin{tabular}{| c | c | c |}
        \hline
        $\; 0 \;$ &           & $\;\;\;\;$ \\
        \hline
                  & $\; 2 \;$ &            \\
        \hline
                  &           &            \\
        \hline
    \end{tabular}
  \end{minipage}
  \hspace{0.5cm}
  \begin{minipage}{1in} \centering
    \begin{tabular}{|c | c | c |}
        \hline
        $\;\;\;\;$ &         &         \\
        \hline
                   & $\;2\;$ &         \\
        \hline
                   &         & $\;1\;$ \\
        \hline
    \end{tabular}
  \end{minipage}
  \hspace{0.5cm}
  \begin{minipage}{1in} \centering
    \begin{tabular}{|l | c | r |}
        \hline
        $\; 0 \;$ &            &         \\
        \hline
                  & $\;\;\;\;$ &         \\
        \hline
                  &            & $\;1\;$ \\
        \hline
    \end{tabular}
  \end{minipage}
  \hspace{0.5cm}
   \begin{minipage}{1in} \centering
    \begin{tabular}{|l | c | r |}
        \hline
        $\; 0 \;$ & $\; 1 \;$ & $\; 2 \;$ \\
        \hline
        $\; 1 \;$ & $\; 2 \;$ & $\; 0 \;$ \\
        \hline
        $\; 2 \;$ & $\; 0 \;$ & $\; 1 \;$ \\
        \hline
    \end{tabular}
  \end{minipage}
 \begin{flushleft}~~~~~~~~ $C_1$ ~~~~~~~~~~~~~~~~~~~~~~~~~  $C_2$ ~~~~~~~~~~~~~~~~~~~~~~~~ $C_3$ ~~~~~~~~~~~~~~~~~~~~~~~~~
 $L$
 \end{flushleft}
\end{table}

We can easily verify that all the partial Latin squares $C_1, C_2,
C_3$ are critical sets. They can be extended uniquely to the full
Latin square in $L$. This unique completion property does not hold
any more if any entry of any partial Latin square $C_1,
C_2, C_3$ is deleted. \\

Let $S$ be the union of the three critical sets $C_1, C_2, C_3$.
Then $S = \{(1,1,1), (2,2,3), (3,3,2)\}$. We distribute a triple to
a participant as a share. Any two participants can recover the full
Latin square. So we have a (2, 3) threshold scheme.\\

\noindent (2) The above simple example can be extended to the
following general case. Let $C_1, C_2, C_3, \ldots, C_n$ be the
critical sets of a given Latin square of size $s_1, s_2, \ldots,
s_n$. Each $C_i$ consists of a set of triples as follows:

\begin{center}
$C_1 = \{(x_{11}, y_{11}, k_{11}), \ldots, (x_{1s_1}, y_{1s_1}, k_{1s_1})\}$ \\
$C_2 = \{(x_{21}, y_{21}, k_{21}), \ldots, (x_{2s_2}, y_{2s_2}, k_{2s_2})\}$ \\
$\ldots \;\;\;\; \ldots \;\;\;\; \ldots$\\
$C_n = \{(x_{n1}, y_{n1}, k_{n1}), \ldots, (x_{ns_n}, y_{ns_n}, k_{ns_n})\}$\\
\end{center}

A triple $(x_{ij}, y_{ij}, k_{ij})$ is interpreted as follow:
$x_{ij}$ is the row of the $j$th element in $C_i$, $y_{ij}$ is the
column of the $j$th element in $C_i$, and $k_{ij}$ is the symbol of
the $j$th element in $C_i$.\\

In general, we make $S$ as a union of some critical sets of a given
Latin square $L$ which represents a secret. Then, the dealer
distributes a share in $S$, in this case a triple of the Latin
square, to each participant. Whenever, a group of participants joins
together to form a critical set, the original Latin square, and
hence the secret can be recovered.\\

Chaudhry, Ghodosi, and Seberry \cite{CGS98} proposed a perfect
secret sharing scheme based on Room squares. This can be applied to
Latin square. The idea is to generate shares randomly for all the
participants with the exception of the last participant, whose
shares will be determined by the shares of other participants and
the critical set in such a way that all the shares when summing up
will be equal to the value of the critical set. Modular arithmetic
are done here. \\

\noindent \underline{Example}: \\

\noindent Let $C = \{(0, 0, 0), (1, 1, 1)\}$ be the critical set of
the Latin square $L$ as Tab.~\ref{Tab_Example}. \\
$L = \{(0, 0, 0), (0, 1, 2), (0, 2, 1); (1, 0, 2), (1, 1, 1), (1, 2,
0); (2, 0, 1), (2, 1, 0), (2, 2, 2)\}.$

\begin{table}[ht]
\caption{Calculation of the share for the last
participant.\label{Tab_Example}}
  \begin{minipage}{1in} \centering
    \begin{tabular}{| c | c | c |}
        \hline
        $\; 0 \;$ &           & $\;\;\;\;$ \\
        \hline
                  & $\; 1 \;$ &            \\
        \hline
                  &           &            \\
        \hline
    \end{tabular}
  \end{minipage}
  \hspace{0.5cm}
   \begin{minipage}{1in} \centering
    \begin{tabular}{|l | c | r |}
        \hline
        $\; 0 \;$ & $\; 2 \;$ & $\; 1 \;$ \\
        \hline
        $\; 2 \;$ & $\; 1 \;$ & $\; 0 \;$ \\
        \hline
        $\; 1 \;$ & $\; 0 \;$ & $\; 2 \;$ \\
        \hline
    \end{tabular}
  \end{minipage}
 \begin{flushleft}~~~~~~~~~ $C$ ~~~~~~~~~~~~~~~~~~~~~~~ $L$
 \end{flushleft}
\end{table}

Let $\{P_1, P_2, P_3\}$ be an authorized set over $C$. Suppose we
generate the following random shares $S_1, S_2$ for $P_1$ and $P_2$
as: $S_1 = \{(0, 1, 2), (2, 0, 0)\}$ and $S_2 = \{(1, 2, 1), (0, 2,
1)\}$. Then share $S_3$ for $P_3$ will be calculated as:\\

\noindent  $S_3 = \{(0-(0+1), 0-(1+2), 0-(2+1)), (1-(2+0), 1-(0+2),
1-(0+1))\} = \{(2, 0, 0), (2, 2, 0)\}.$ \\

All arithmetic are done in $mod~3$. It can be easily verified that
$P_1, P_2, P_3$ can recover the critical set when they pool their
shares together. If any participant is missing, it makes the
unauthorized set contain nothing more than any outsider. \\

To summarize, there are reasons why we want to apply critical sets
to secret sharing scheme:

\begin{enumerate}[1)]
\item Since a critical set can always be extended to a full Latin
square uniquely, it would be more efficient to distribute shares of
a critical set rather than a full Latin square.

\item A $(t+1, n)$ threshold scheme or multilevel scheme can be implemented through critical
sets, as discussed in Chaudhry, Ghodosi, and Seberry \cite{CGS98}.
\end{enumerate}

\section{Limitations of Latin square based secret sharing schemes}

Many researches have been done since the original secret sharing
ideas of Shamir \cite{SHA79} and Blakley \cite{BLAK79} in 1979.
Latin square was suggested as a good candidate being used in secret
sharing schemes. However, there are certain limitations as discussed
below.\\

\noindent 1) By just distributing shares of a critical set to
participants, partial information will be available to any
unauthorized group. That means there is a good chance for any
unauthorized group to figure out the remaining shares by trial and
error method. So, the scheme proposed by Cooper, Donovan, Seberry
\cite{CDS94} is not perfect. \\

\noindent 2) The scheme proposed by Chaudhry, Ghodosi, Seberry
\cite{CGS98} is not flexible if there is only one authorized set. In
this case it is just a secret splitting scheme. If more than one
authorized set exists, the secret sharing scheme is not ideal. Each
participant needs to have different share for different authorized
set he/she belongs to. \\

\noindent 3) As we know, distributing shares of a critical set
instead of a Latin square is definitely more desirable. However,
there are two issues need to be considered:

\begin{enumerate}[(a)]
\item Even getting all the shares about a critical set, it may not be
easy to get back the original Latin square, the shared secret. In
order to speed up the recovering process, we should use a strong
critical set.

\item However, if the participants of an authorized group join
together, it will be much easier for them to figure out the
shared secret if the chosen critical set is a strong one. \\
\end{enumerate}

\noindent 4) The knowledge about the critical sets of Latin squares,
especially of large order (say 10), is very limited. There are
critical sets of different size. It is very difficult to verify or
find a critical set. These hinder the implementation of various
secret sharing schemes based on critical sets.

\begin{enumerate}[(a)]
\item Control: Let $S$ be a collection of critical sets $C_1, C_2,
C_3$ of Latin square $L$. We would like to design a secret sharing
scheme such that any authorized set of participants can recover
$C_1$ or $C_2$ or $C_3$. But there is a possibility that $S$
contains another critical set $C_4$. If individuals of any
unauthorized set (in the sense that they cannot recover $C_1, C_2$
or $C_3$) can pool their shares to form $C_4$, then they can recover
$L$. Hence some careful controls need to be taken especially given
the condition that critical set of large order Latin square is
difficult to find or verify.

\item Implementation: It would not be so flexible and easy to set up
a verifiable sharing scheme, a proactive sharing scheme, or a $(t+1,
n)$ threshold scheme just by using a Latin square or some of its
critical sets to represent the secret especially when we choose a
Latin square of order greater than 10 due to the limited knowledge
about its critical set.
\end{enumerate}

\section{Apply hash function to Latin square based secret sharing schemes}

Zheng, Hardjono, and Seberry \cite{ZHS94} discuss how to reuse
shares in a secret sharing scheme by using universal hash function.
In this Section,  we'll show how to use general hash function
properties including herding, and Nostradamus attacks \cite{KK06} to
design and improve Latin square based secret sharing schemes.

\subsection{Store Latin square in a hash\label{Sec:StoreHash}}

If we want to use the hash to store a fixed secret, for example, a
Latin square of order 10, we need to store 81 numbers (since the
last row and last column are not necessary). Four bits can be used
to store a number, so we need 324 bits. In this case, we can choose
SHA-384 or SHA-512 to fulfill the requirements easily.\\

If we need to use SHA-256, we can proceed in the following way. 10
bits can be used to represent 3 numbers. So, we first use 250 bits
to represent 75 numbers and then the next 4 bits to represent a
single number. Altogether, we can store 76 numbers. We fix the
partial Latin square in the following format.\\

\begin{table}[ht]
\begin{center}
\caption{Use 10-bit to represent 3 numbers in Latin square of order
10.\label{Tab_Example_2}}
\begin{tabular}{| c | c | c | c | c | c | c | c | c | c |}
    \hline
    $\;\; \times \;\;$ & $\;\; \times \;\;$ & $\;\; \times \;\;$ & $\;\; \times \;\;$ & $\;\; \times \;\;$ & $\;\; \times \;\;$ & $\;\; \times \;\;$ & $\;\; \times \;\;$ & $\;\; \times \;\;$ & $\;\; \;\;\; \;\;$\\
    \hline
    $ \times $ & $ \times $ & $ \times $ & $ \times $ & $ \times $ & $ \times $ & $ \times $ & $ \times $ & $ \times $ & $ \; $\\
    \hline
    $ \times $ & $ \times $ & $ \times $ & $ \times $ & $ \times $ & $ \times $ & $ \times $ & $ \times $ & $ \times $ & $ \; $\\
    \hline
    $ \times $ & $ \times $ & $ \times $ & $ \times $ & $ \times $ & $ \times $ & $ \times $ & $ \times $ & $ \times $ & $ \; $\\
    \hline
    $ \times $ & $ \times $ & $ \times $ & $ \times $ & $ \times $ & $ \times $ & $ \times $ & $ \times $ & $    \;  $ & $ \; $\\
    \hline
    $ \times $ & $ \times $ & $ \times $ & $ \times $ & $ \times $ & $ \times $ & $ \times $ & $ \times $ & $    \;  $ & $ \; $\\
    \hline
    $ \times $ & $ \times $ & $ \times $ & $ \times $ & $ \times $ & $ \times $ & $ \times $ & $ \times $ & $    \;  $ & $ \; $\\
    \hline
    $ \times $ & $ \times $ & $ \times $ & $ \times $ & $ \times $ & $ \times $ & $ \times $ & $ \times $ & $    \;  $ & $ \; $\\
    \hline
    $ \times $ & $ \times $ & $ \times $ & $ \times $ & $ \times $ & $ \times $ & $ \times $ & $ \times $ & $    \;  $ & $ \; $\\
    \hline
    $ \; $     & $ \; $     & $ \; $     & $ \; $     & $ \; $     & $ \; $     & $ \; $     & $ \; $     & $    \;  $ & $ \; $\\
    \hline
    \end{tabular}
    \end{center}
\end{table}

We choose a Latin square of order 10 that can be recovered uniquely
by removing the entries as shown in Tab. \ref{Tab_Example_2}. The
tradeoff here is that a small percentage of Latin squares of order
10 can not be recovered uniquely and hence cannot be chosen as
secret. \\

We want to recover the number in (4, 8), (5, 8), (6, 8), (7, 8), (8,
8) in the following way. Pick any row between 4th and 8th. If $a$
and $b$ are the number missed in row $I \;(4 \leq I \leq 8)$ and
$a(b)$ is in the 8th column, we can fill in $b(a)$ in the $(I, 8)$
cell. If we can recover (4, 8), (5, 8), (6, 8), (7, 8), and (8, 8)
in this way, we can recover the original Latin square uniquely.\\

Unused bits can be filled in randomly. The above are just simple
examples to demonstrate how to use hash to represent fixed secret.



\subsection{Set up an ideal perfect $(t + 1, n)$ threshold scheme \label{Sec:Perfect}}

Let's continue with Section~\ref{Sec:StoreHash} and suppose the
secret is the hash of a (partial) Latin square. Let's consider how
to apply a hash function $f$ to set up a $(t + 1, n)$ threshold
secret sharing scheme. The approach we take is based on herding hash
technique. \\

First we randomly generate a share of more or less the same size as
that of the hash to each participant. Then, we set up different
authorized subsets so that each subset consists of $(t + 1)$ or more
distinct participants. \\

Let $N$ be the size of the access structure, i.e., the total number
of all authorized subsets.

$$N = C(n, t+1) + C(n, t+2) + \ldots + C(n, n),$$

\noindent where $C(n, t) = (n!)/(t!(n-t)!)$ is the combination
function. That means we need to have $N$ messages for these $N$
authorized subsets. There is a one-to-one correspondence between
messages and authorized subsets. \\

Each participant holds a share and any combination of the shares of
an authorized subset will generate one of these $N$ messages. The
next step is to herd the hashes of these $N$ messages into the final
hash as the Nostradamus attack by setting up the linking messages.
\\

\begin{figure}
\centering
\includegraphics[height=2.5cm, width=8cm]{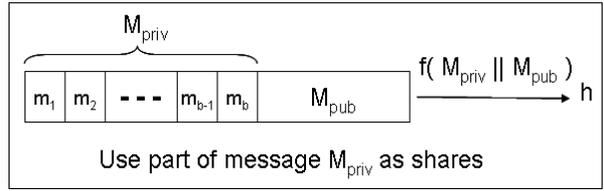}
\caption{Message $M$ and sub-messages, i.e., shares $m_i$.}
\label{fig:MSPLIT}
\end{figure}

Suppose an authorized set consists of participants $P_1, P_2,
\ldots, P_b$ and their shares are sub-messages $m_1, m_2, \ldots,
m_b$. When they join together, they can form $M_{priv} = m_1 ||
\ldots || m_b$ and find the corresponding linking message $M_{pub}$,
as shown in Fig.~\ref{fig:MSPLIT}. Then they can recover the secret
$h$ by applying the hash function $f$ to $M_{priv}||M_{pub}$, i.e.,
$f(M_{priv}||M_{pub}) = h$.

In the Nostradamus attack, we don't know what will happen, so we
need to

\begin{enumerate}[a)]
\item build a huge diamond structure leading to a final hash $h$;
\item find a linking block after the result is known.
\end{enumerate}

In our case, the above steps are not necessary since we know the
hashes of these $N$ messages. This greatly reduces the effort.\\

For any message $M_{priv}$ obtained by combining the shares of the
participants in an authorized subset, there is a corresponding
message $M_{pub}$ in the diamond structure. Linking these two
messages can reach the final hash of the diamond structure. So, we
have a $(t+1, n)$ threshold scheme based on herding hash functions
technique. The linking messages are stored in a public place which
can be accessed by any participant. When any group of $t + 1$ or
more participants join together, they can look for the corresponding
linking message and plus their shares to recover the secret.\\

Properties of the proposed scheme include:

\noindent a) Perfect: One of the basic properties of a cryptographic
hash function is its randomness. Based on the message, we cannot
figure out any information about the hash. This avoids revealing
partial information to any participant. When all participants join
together, they can recover the secret by applying the hash function
$f$ to the message $M = M_{priv}||M_{pub}$. In order to maintain the
security level, the length of each share should be at least as long
as the hash. On the other hand, increasing the length of the share
does not increase the security level. So, we would like to have each
share to be generated randomly and of length more or less the same
as the hash. This will be the case if the message was generated
randomly. This provides a perfect sharing scheme because even one
participant is missing, the share cannot be recovered and no
information about the secret is leaking out. \\

\noindent b) Ideal: The scheme is ideal since each participant holds
one share which has the same size of the hash. \\

\noindent c) Fast recovery of secret: The calculation of hash
function is fast, this can assure that the partial Latin square and
hence the full Latin square can be recovered quickly.\\

\noindent d) Avoid of critical sets: Under the new scheme, looking
for critical sets of large size can be avoided. This makes it more
efficient and better controlled as discussed above.\\

\noindent e) Application of minimal authorized subset: We provide a
complete description here. But, as we shall see in the example, we
can speed up the whole process by considering the minimal authorized
subset only.\\

\noindent f) General access  structure: As we shall see in the
following example, this approach can be extended to general access
structure.\\

\noindent \underline{Example}:\\

\begin{figure}
\centering
\includegraphics[height=4.5cm, width=8cm]{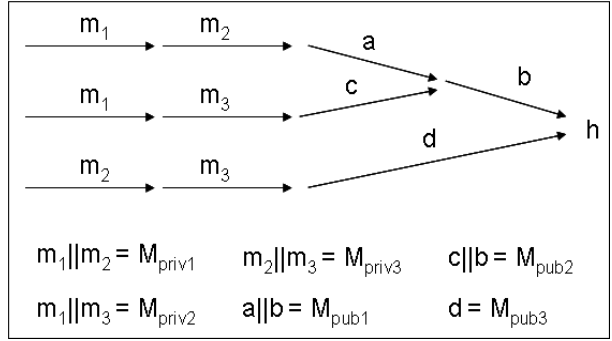}
\caption{A (2, 3) threshold scheme example.}
\label{fig:SimpleDiamond}
\end{figure}

A (2, 3) threshold scheme. Let $m_1, m_2$, and $m_3$ be shares of
participants $P_1, P_2$, and $P_3$, respectively. Then, the access
structure consists of four authorized subsets, also shown in
Fig.~\ref{fig:SimpleDiamond}. ${M_{pub1}, M_{pub2}, M_{pub3},
M_{pub4}}$ will be the linking messages stored in the public area.

\begin{enumerate}[a)]
\item $\{P_1, P_2\} \hspace{3.5em} m_1||m_2||M_{pub1}$
\item $\{P_1, P_3\} \hspace{3.5em} m_1||m_3||M_{pub2}$
\item $\{P_2, P_3\} \hspace{3.5em} m_2||m_3||M_{pub3}$
\item $\{P_1, P_2, P_3\} \hspace{2.0em} m_1||m_2||m_3||M_{pub4}$
\end{enumerate}

While it would be straight forward to set up the access structure
with all the authorized groups, it would be more efficient if we
only consider the minimal authorized subset  of the access 
structure. In this case, we can skip $m_1||m_2||m_3||M_{pub4}$.\\

Suppose we know $P_2$, $P_3$ are family members or good friends, we
don't want them to recover the secret. Then, a general $(2,3)$
threshold scheme doesn't work. For our case, we can just simply skip
the setup of $m_2||m_3||M_{pub3}$. \\

It is easy to show that this method is good for any general access
structure. \\

\subsection{Set up a verifiable scheme \label{Sec:Verifiable}}

A cryptographic hash function has an application as message
authentication code to certify that original message was not
altered. We can apply this idea to secret sharing scheme so that any
dishonest participant who does not return the original share will be
found by the dealer. On the order hand, the participants can verify
whether the dealer really sends out consistent shares for them to
keep. So, let us modify \ref{Sec:Perfect} approach for an
implementation of a verifiable secret sharing scheme.\\

Let $f, g$ be cryptographic hash functions. Let $M$ be a message
such that $f(M) = s$ where $s$ is the shared secret. The dealer
breaks $M$ into different sub-messages $m_1, m_2, \ldots,$ and
distributes each share to each participant and then publishes the
hashes (by hash function $g$) of each share as commitments:
 $g_1, g_2, \ldots,$ as in Feldman's case. \\

Participant $i$ verifies his/her share by checking if $g(m_i) = g_i$
holds. If all participants confirm that taking his/her share as
input to the hash function $g$, he/she gets the hash value equals to
one of the commitments published by the dealer, we conclude the
dealer sends out consistent shares. Likewise, when the participants
return their shares, the dealer can verify in the same way.\\

As we can see from the above, we use two hash functions $g$ and $f$.
Hash function $g$ is used to make the scheme as an verifiable secret
sharing scheme. Hash function $f$ is used to recover the shared
secret: $f(M)$. Participant $i$ can fool the party if he/she can
find $m_i'$ such that $g(m_i) = g(m_i') = g_j$. If $g$ is second
preimage resistant, this is difficult to achieve and the scheme is
safe.

\section{Conclusion}

In this paper, we use cryptographic hash functions to improve the
security and performance of secret sharing schemes based on a Latin
square or its critical sets. We can store a partial Latin square in
a hash for a fast retrieval of the shared secret; we can set up an
ideal perfect $(t+1, n)$ threshold secret sharing scheme with easily
extendable to have verifiable, proactive, hierarchical properties.
This can also apply to any general access structure.

\subsubsection*{Acknowledgments.} Authors would like to thank Prof.
Michael Anshel for his valuable discussions. We also want to thank
Prof. Joseph Vaisman for getting us many useful references.

\bibliographystyle{plain}
\bibliography{Latin_square_SSS_v2}

\end{document}